\documentclass[useAMS,usenatbib,usegraphicx]{mn2e}

\title[Formation of star clusters]{Self-consistent simulations of star cluster formation  
from gas clouds under the influence of galaxy-scale tidal fields} 

\author[J. R. Hurley and K. Bekki]
{Jarrod R. Hurley$^{1}$\thanks{E-mail: jhurley@swin.edu.au (JRH)} 
and
Kenji Bekki$^{2}$\thanks{E-mail: bekki@phys.unsw.edu.au (KB)} \\
$^{1}$Centre for Astrophysics and Supercomputing, Swinburne University of Technology, P.O. Box 218, VIC 3122, Australia \\
$^{2}$School of Physics, University of New South Wales, 
Sydney 2052, NSW, Australia}

\begin{document}

\date{Accepted 2008 Month xx. Received 2008 Month xx; in original form 2008 March xx} 

\pagerange{\pageref{firstpage}--\pageref{lastpage}} \pubyear{2008}

\maketitle

\label{firstpage}

\begin{abstract}
We present the first results of a project aimed at following the formation and long-term 
dynamical evolution of star clusters within the potential of a host galaxy. 
Here we focus on a model evolved within a simplified potential 
representing the Large Magellanic Cloud. 
This demonstrates for the first time the self-consistent formation of a bound star cluster from 
a giant molecular cloud. 
The model cluster reproduces the density profiles and structural characteristics 
of observed star clusters. 
\end{abstract}


\section{Introduction}
\label{s:intro}

Star clusters are vital components of galaxies. 
Indeed, extracting information from the massive globular clusters (GCs) 
of our Galaxy provides fundamental information on the epoch of galaxy 
formation. 
Furthermore, systems of extragalactic GCs are used as key determinants for tracing 
the dynamical, chemical and gaseous evolution of their host galaxies 
(see Brodie \& Strader 2006 for a review). 
The smaller open clusters also have a role to play as well, as understanding 
their destruction within in the Galactic disk impacts the growing field 
of Galactic archaeology (Freeman \& Bland-Hawthorn 2002), for example.  

Efforts to model the evolution of individual star clusters 
have made great advances in the last decade. 
On the hardware front the advent of special-purpose GRAPE processors 
for calculating gravitational forces, with the latest incarnation being 
the GRAPE-6 (Makino 2002), have allowed $N$-body models of up 
to $N \sim 100\,000$ stars to be completed in a reasonable timeframe 
(Baumgardt \& Makino 2003). 
The next generation of GRAPE (available in 2008), the use of massively-parallel 
supercomputers, and even graphics processing technology, will push 
this $N$ continually higher towards the realm of direct GC models. 
Complementing this is the push to make the models as realistic as possible 
by including algorithms to deal with processes such as stellar and binary evolution 
in concert with the treatment of the gravitational interactions of the stars. 
This is the case for both $N$-body (Hurley et al. 2005) and 
statistical (Fregeau \& Rasio 2007; Giersz, Heggie \& Hurley 2008) techniques. 

On a grander scale simulations of galaxy formation which identify the formation 
sites of star clusters have been 
performed (Bekki \& Chiba 2005). 
Furthermore, the formation of stars within turbulent giant molecular clouds (GMCs) 
based at these formation sites can then be followed with smooth-particle 
hydrodynamics (SPH) simulations (Bekki \& Chiba 2007). 
Our aim is to interface these galactic-scale simulations of star cluster formation 
with the latest modeling techniques for following the long-term evolution 
of star clusters. 
The immediate benefits will be: a) the first self-consistent simulations of star 
cluster evolution from formation through to destruction, and b) 
a non-simplistic picture of how GC systems evolve with time and how 
this impacts the interpretation of observed extragalactic GC systems. 
This latter advance in particular will be important for understanding the 
connection between extragalactic GCs and galaxy formation by injecting much 
needed numerical models in to what has become a data-dominated field 
(Brodie \& Strader 2006). 
We will also be able to explore the origin of observed features of star cluster systems, 
such as the age and parameter distributions of the 
Large Magellanic Cloud (LMC) 
clusters (Mackey \& Gilmore 2003). 

In this letter we provide an overview of the modeling process that 
will be used to achieve our goals. 
We illustrate this process by describing the evolution of a prototype model 
which demonstrates a working interface between the cluster formation and 
long-term evolution codes. 
It also entails the first simulation of bound cluster formation from a gas cloud 
within an external galactic potential. 
We then discuss future improvements to the model and details of how 
this fits in to our plans for a full investigation of the long-term evolution 
of galactic and extragalactic star cluster systems.

\section{A Two-fold Modelling Procedure}
\label{s:models}

The present model is two-fold as follows.
We first determine the spatial distributions of stars that are formed
within GMCs gravitationally influenced by their host galaxies.
In this first step, we investigate short-term star formation
processes in GMCs based on GRAPE-SPH simulations of turbulent GMCs. 
Then we investigate the long-term dynamical evolution of collisional
stellar systems, incorporating processes such as 
two-body relaxation and mass-loss from stars, 
for initial distributions of stars derived in the first step.
In this second step, we emphasize that we are taking 
{\it dynamically unrelaxed stellar systems} formed from GMCs 
and placing these within an $N$-body code 
for following the evolution of very young
SCs gravitationally influenced by the tidal field of their
host galaxies.

Previous work along these lines that bears mention starts with the 
simulations of galaxy formation reported by van Albada (1982, 
and see associated references within) and McGlynn (1984). 
These dealt with similar physical processes -- the evolution of an 
initially irregular mass distribution -- although the context and approach 
were quite different. 
Importantly these were collision-free simulations performed without the 
influence of an external potential. 
Aarseth, Lin \& Papaloizou (1988) examined the collapse of proto-globular 
clusters through $N$-body models. 
They took an initial distribution of fragments 
(representing low-mass pre-stellar subcondensations) 
and documented the ensuing gravitational relaxation phase leading to the 
creation of a core-halo density structure. 
These models also did not include an external potential. 
Furthermore, each of these studies did not model the fragmentation process 
leading to the formation of gas clumps or the subsequent creation of stars 
from these clumps. 
These prior studies do however provide a good basis to which we can compare 
the results of our models.

\subsection{Turbulent GMCs} 

Radial mass density distributions, $\rho (r)$, of GMCs with the sizes, $r_{\rm g}$, and
masses, $m_{\rm g}$, are assumed to have homogeneous spherical distributions.
Here we set up the initial velocity fields due to turbulent flows within GMCs in the same way 
as Mac Low et al. (1998). 
We therefore assume that a turbulent velocity field  within a GMC is a Gaussian random
field with a power spectrum for $0 \le {\bf k} \le k_{\rm max}$ as follows: 
\begin{equation}
P(k) = P_0 k^{\alpha} \, , 
\end{equation}
where
$\alpha$ is set to be 2.0 for most models and
$P_0$ is a parameter controlling the total kinematical energy
due to the turbulent flow in the GMC.
We mainly use $k_{\rm max}=8$ in our models, however, 
the final dynamical and chemical properties
of the simulated GCs  do not depend strongly on $ k_{\rm max}$.

The virial ratio, $t_{\rm v}$, is a free parameter described as: 
\begin{equation}
t_{\rm v} = 2T_{\rm k}/W = f(P_0) \, , 
\end{equation}
where $T_{\rm k}$ and $W$ are the total kinematical energy and the 
absolute magnitude of the total potential energy for a GMC, respectively.
As shown in equation (2),
$t_{\rm v}$ ($0 \le t_{\rm v}\le 1$) is determined by $P_0$ and
thus can control the initial random kinematical
energy of gas particles in the present study.
We typically take $t_{\rm v} = 0.25$ in our models. 
Since an isothermal equation of state
is suggested to be appropriate for star-forming
interstellar clouds of molecular gas  (e.g., Mac Low et al. 1998; Klessen, Heitsch \& Mac Low 2000),
we adopt the equation with the initial temperature of 10 K.

\subsection{Star formation and stellar feedback effects}

Gas particles with initial masses
of $m_{i}$ are assumed to be converted into new stellar particles 
if (i) local dynamical time scales are shorter than the local sound
crossing time, and (ii) local gas densities exceed a threshold gas
density, ${\rho}_{\rm th}$, of star formation (e.g., Nakasato, Mori \& Nomoto 2000).
Since the gravitational softening length, ${\epsilon}$, for gas particles
of GMCs in a model is fixed during the evolution of the GMC
($\epsilon \sim 10^{-2} r_{\rm g}$),
there is a maximum density, ${\rho}_{\rm max}$, which the 
gas densities of individual SPH particles, ${\rho}_{i}$, can not exceed
in the adopted GRAPE-SPH method:
${\rho}_{\rm max} \sim N_{\rm n} m_{i}/{{\epsilon}_{\rm g}}^{3}$,
where $ N_{\rm n}$ is the total  number of ``neighbor particles'' surrounding
an $i$-th SPH particle.
Here ${\rho}_{\rm max}$ is  estimated to be of the order of  $10^3$ atom cm$^{-3}$
for  models with $m_{\rm g} \sim 10^6 {\rm M}_{\odot}$.
We thus assume that ${\rho}_{\rm th}={\rho}_{\rm max}$ in the present study.
It should be noted here that the above ${\rho}_{\rm max}$ is significantly lower 
than the threshold gas density ($\sim 10^5$ atom cm$^{-3}$) of {\it individual stars} 
suggested by Elmegreen (2004).

\subsection{External gravitational potential} 

As star clusters evolve they lose stars to their host galaxy at a rate that depends on
the strength of the galactic potential and the orbit within this potential.
Our plan is to improve the treatment of the external 
gravitational potential used in simulations of star cluster evolution 
by using the results of the galaxy-scale calculations to provide 
a `live' model of the potential. 
However, to start with we adopt a smooth and static potential that 
is the current standard for $N$-body star cluster models. 
We tailor this to an external gravitational potential reasonable for the LMC 
in our initial study as one of our goals is to compare our results with the 
observed properties of young star clusters in the LMC. 

Star-forming GMCs are thus assumed to be gravitationally influenced by
the LMC represented by a point-mass, $M_{\rm gal}$, of $0.9 \times 10^{10} {\rm M}_{\odot}$. 
Within this simplified potential the 
GMCs are assumed to have circular velocities determined
by their locations and the mass of the LMC.

\subsection{Dynamical evolution of SCs just after their birth}

The long-term dynamical evolution of the young SCs emerging from the GMCs 
is then followed using the {\tt NBODY4} code (Aarseth 1999, 2003). 
This code employs the fourth-order Hermite integration scheme, without 
softening, and is designed to exploit the fast evaluation of the gravitational force 
and its time derivative by the GRAPE-6. 
The use of the GRAPE-6 allows models of up to $N = 100\,000$ to be completed 
although for $N < 10\,000$ it is possible to make 
progress even when performing the full force calculation on the host computer. 
{\tt NBODY4} includes algorithms for stellar and binary evolution as described in 
Hurley et al. (2001). 
It allows for the full range of possible interactions within binary stars as well 
as dealing directly with the effects of close dynamical encounters: 
perturbations to binary orbits, collisions and mergers, formation of three- and 
four-body subsystems, exchange interactions, tidal capture and binary disruption. 
The tidal field of the host galaxy is modelled as described in Sec. 2.4 by placing 
the model cluster on a circular orbit at a specified radial distance, $R_{\rm gc}$, 
from the centre of a point-mass galaxy. 
As such {\tt NBODY4} facilitates an investigation of the combined effects of 
internal (stellar, binary and dynamical) and external (galactic tide) 
influences on the evolution history of a star cluster. 

The typical starting point for models of star cluster evolution is to assume 
that the star formation process is complete. 
An initial model is designed by first drawing the positions of the stars from 
some density distribution: 
the $n = 5$ polytrope proposed by Plummer (1911) and the family of 
King models (King 1962, 1966) derived from observations of mature star 
clusters are common choices, the former primarily because of its mathematical 
simplicity. 
Next the stellar velocities are determined by assuming the model cluster is 
in dynamical, or virial, equilibrium, i.e. 
\begin{equation}
Q_{\rm v} = T_{\rm k} / \left( W - 2 E_{\rm t} \right) = 0.5 \, 
\end{equation} 
where we have switched to the virial ratio definition used in star cluster simulations 
and $E_{\rm t}$ represents the energy contribution from external forces 
(see Fukushige \& Heggie 1995; Aarseth 2003, p.~11). 
The masses of the stars are assumed to be either equal or drawn from 
an initial mass function based on observations of field stars.  
In our new method we do away with these assumptions and feed the results 
of the star cluster formation calculations directly in to the {\tt NBODY4}  code. 
That is, the masses, positions and velocities of the newly formed stars. 
In this way we can ascertain if a bound cluster eventuates, 
while following the evolution of the density and velocity profiles of the young cluster.

\section{Results of a prototype model}
\label{s:results}

\begin{figure}
\includegraphics[width=84mm]{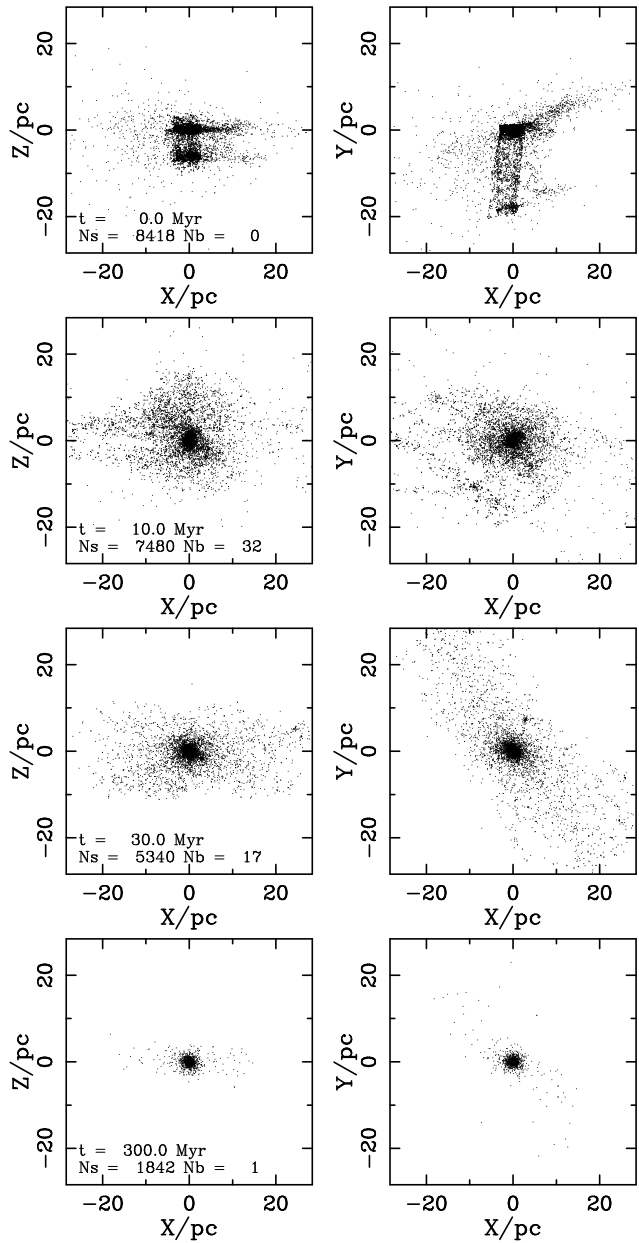}
\caption{
Spatial distributions in the XZ- and XY-planes for the prototype 
model at ages of 
0, 10, 30 and $300\,$Myr. 
The number of single stars, $N_{\rm s}$, and binaries, $N_{\rm b}$, 
bound to the cluster at each age are denoted in the figure. 
\label{f:fig1}}
\end{figure}

In this letter we highlight our procedure by describing the evolution of 
a proto-cluster of $8\,420$ equal-mass stars emerging from a GMC. 
We set $m_{\rm g} = 10^4 M_\odot$ and $r_{\rm g} = 10\,$pc 
for the GMC and take the number of gas particles to be $2 \times 10^4$. 
Thus the mass of the stars formed is $0.5 M_\odot$ per star. 
The timescale of star formation in this step is $\sim 3 \times 10^6\,$yr 
and the star formation efficiency is 42\%. 
Full details of the formation process will be supplied in an upcoming paper. 
The spatial distribution of the proto-cluster is shown in the top panels 
of Figure~\ref{f:fig1}. 
This is the initial model for the {\tt NBODY4} step of the evolution 
and as such is given an age of $0\,$Myr. 
In this initial model there are no gas particles remaining but 
residual gas can be communicated to the $N$-body code, and its 
effect accounted for, in any future models. 
The stars in the initial model do not constitute a bound system 
(the total energy is positive) and are definitely not in virial 
equilibrium: the ratio of kinetic to potential energy is 1.84 
(this is $Q_{\rm v}$ from eq. 3 with $E_{\rm t} = 0$). 
The half-mass radius, $r_{\rm h}$, is $5\,$pc, the velocity dispersion 
is $2.8 \, {\rm km} \, {\rm s}^{-1}$, and the corresponding crossing-time 
for this initial model is approximately $7\,$Myr. 

The cluster was placed on a circular orbit within our simplified 
LMC potential at a distance of $R_{\rm gc} = 1.5\,$kpc from the galaxy centre. 
This matches the location of the parent GMC and corresponds to the 
scale length of the stellar disk of the LMC. 
The choice of tidal field gives an initial tidal radius of $8\,$pc. 
We note that as the model cluster evolves, stars are denoted as having escaped from 
the cluster (and are removed from the simulation) 
when their distance from the cluster centre exceeds twice the tidal radius. 
This prototype model was evolved to an age of $550\,$Myr using a single processor 
of the Swinburne supercomputer. 
At this point a bound cluster comprising 10\% of the original stars remains. 

\begin{figure}
\includegraphics[width=84mm]{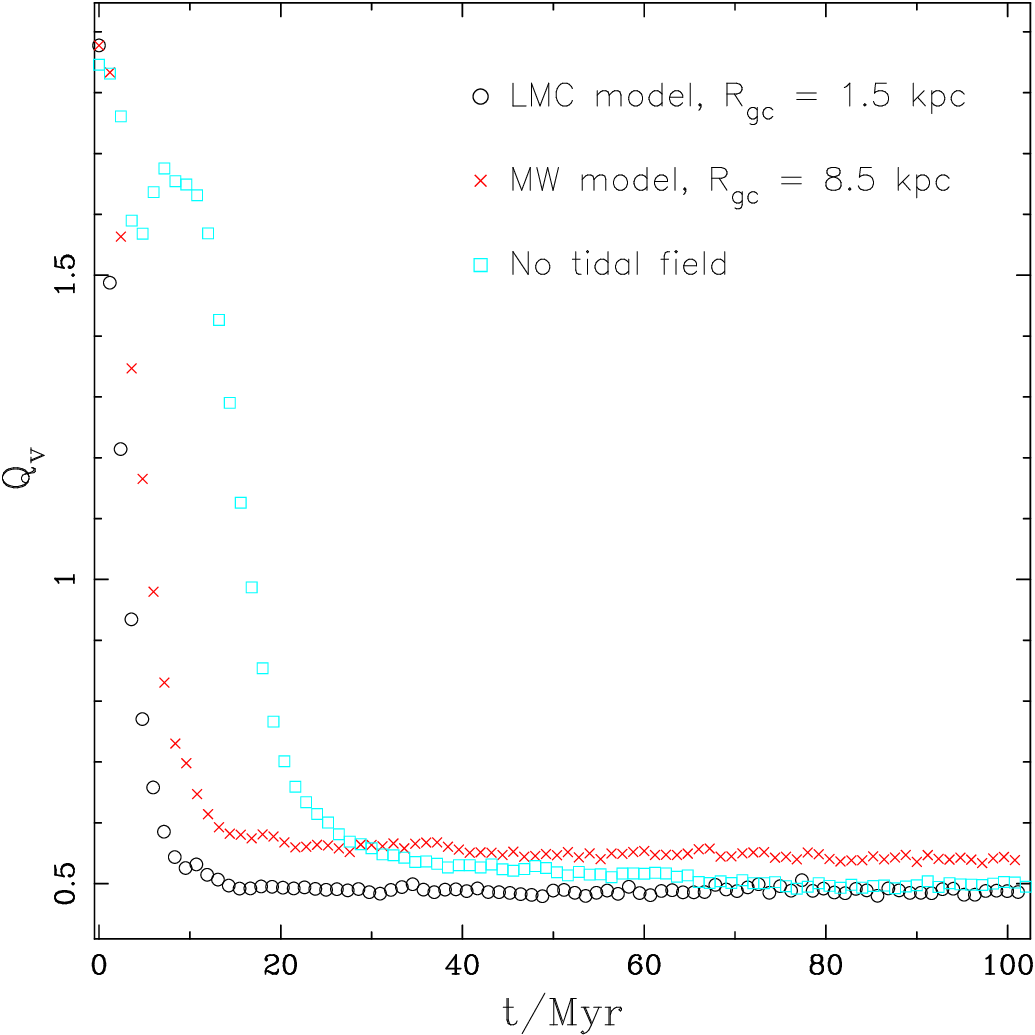}
\caption{
Evolution of the virial ratio for stars in the prototype model (open circles). 
Also shown is the same model but evolved at a distance of $8.5\,$kpc from 
the centre of a point-mass Milky Way galaxy (red x symbols), 
and evolved as an isolated cluster with no external force (cyan open squares). 
\label{f:fig2}}
\end{figure}

The total energy of the system first becomes negative after $5\,$Myr have elapsed. 
Figure~\ref{f:fig1} shows the spatial distribution of the stars at ages of 10, 30 
and $300\,$Myr. 
Already at $10\,$Myr the makings of a star cluster are evident 
and certainly by $300\,$Myr the system has the regular appearance of a tightly 
bound star cluster. 
The virial ratio for the cluster stars (as defined by equation 3) is shown in Figure~\ref{f:fig2} 
for the first $100\,$Myr of evolution. 
We see that the ratio steadily decreases and that by an age of $30\,$Myr the cluster has 
reached approximate virial equilibrium. 
Certainly by an age of $55\,$Myr, 
when a two-body relaxation time (as measured at the half-mass radius) has passed, 
the cluster appears to be dynamically relaxed.  
When the model cluster is $100\,$Myr old the escape rate of stars from the 
cluster is 14/Myr. 
This is down from an average of 80/Myr during the first $55\,$Myr of evolution 
but matches the escape rate from a comparison cluster started in virial equilibrium 
and evolved on the same LMC orbit. 
The model at $100\,$Myr contains $2\,817$ bound stars. 
It has spatial parameters of $r_{\rm t} = 5.6\,$pc, $r_{\rm h} = 1.2\,$pc and 
core-radius, $r_{\rm c} = 0.36\,$pc. 
The half-mass relaxation timescale has decreased to $28\,$Myr and the velocity 
dispersion is $1.5 \, {\rm km} \, {\rm s}^{-1}$. 

For comparison we have evolved the same initial model but 
at a distance of $R_{\rm gc} = 8.5\,$kpc from the centre of a point-mass galaxy 
with $M_{\rm gal} = 9 \times 10^{10} {\rm M}_{\odot}$ 
-- this resembles the orbit of an open cluster residing in the Solar neighborhood 
of the Galactic disk and is commonly referred to as a standard Galactic tidal field 
(Giersz \& Heggie 1997). 
Within a point-mass galaxy the tidal radius for a cluster of mass, $M_{\rm c}$, scales 
as 
\begin{equation}
r_{\rm t} \propto \left( M_{\rm c} / M_{\rm gal} \right)^{1/3} R_{\rm gc} \, . 
\end{equation}
So the Milky Way (MW) model with $M_{\rm gal,MW} = 10 \, M_{\rm gal,LMC}$ and 
$R_{\rm gc,MW} \simeq 6 \, R_{\rm gc,LMC}$ has 
$r_{\rm t,MW} \simeq 3 \, r_{\rm t,LMC}$ and therefore represents a weaker tidal field. 
We see in Figure~\ref{f:fig2} that this leads to the model approaching virial equilibrium 
on a similar timescale to that of the LMC model ($20 - 30\,$Myr). 
To further understand the effect of the tidal field we 
have also evolved the initial model as an isolated cluster 
($E_{\rm t} = 0.0$ at all times). 
This model approaches virial equilibrium on a timescale of $\sim 40\,$Myr. 
Thus we see a weak trend for the virial timescale to increase for models with a 
decreased influence from the external tidal force. 
At the same time, the half-mass relaxation timescale is greater in models with a 
weaker tidal field, owing to larger $N$ and $r_{\rm h}$ at any particular time. 
We note that in equation (3) we have neglected a term that represents the rotation of the 
cluster (the Coriolis force) as this is negligible for clusters in dynamical equilibrium 
(Fukushige \& Heggie 1995). 
However, its absence explains the slight deviation from $Q_{\rm v} = 0.5$ for our relaxed models, 
as does the fact that a cluster can approach a state of dynamical equilibrium but will never 
actually reach it in practice. 
The Milky Way model at an age of  $100\,$Myr contains $6\,724$ stars and has 
$r_{\rm t} = 21\,$pc, $r_{\rm h} = 3.5\,$pc, $r_{\rm c} = 0.36\,$pc 
and a velocity dispersion of $1.4 \, {\rm km} \, {\rm s}^{-1}$. 

\begin{figure}
\includegraphics[width=84mm]{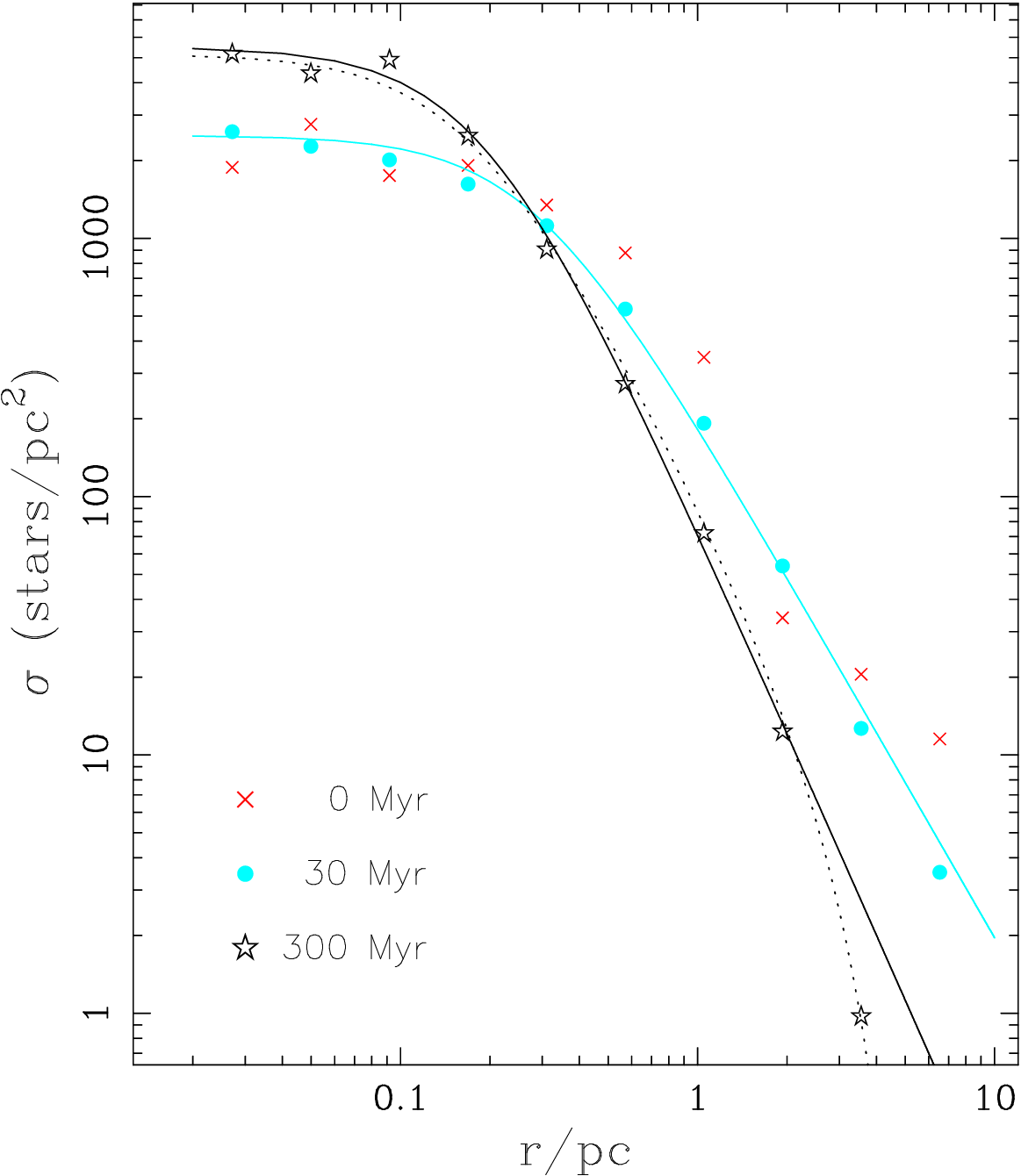}
\caption{
Radial surface density profile of the prototype LMC model at ages of 
0 (red x symbols), 30 (cyan circles) and 300 (open stars) Myr 
%constructed by projecting the data along the Y-axis. 
(projected along the Y-axis). 
Shown also are the best fitting EFF profiles for the $30\,$Myr data 
(cyan solid line: $r_{\rm c} = 0.28\,$pc, 
$\sigma_0 = 2\,500\, {\rm stars} \, {\rm pc}^{-2}$, $\gamma = 2.0$) 
and the $300\,$Myr data 
(black solid line: $r_{\rm c} = 0.16\,$pc, 
$\sigma_0 = 5\,500\, {\rm stars} \, {\rm pc}^{-2}$, $\gamma = 2.6$) 
as well as the best fitting King (1962) model for the $300\,$Myr data 
(dotted line: $r_{\rm c} = 0.16\,$pc, 
$\sigma_0 = 5\,500\, {\rm stars} \, {\rm pc}^{-2}$, $r_{\rm t} = 5.0\,$pc). 
\label{f:fig3}}
\end{figure}

In Figure~\ref{f:fig3} we show the two-dimensional radial density profiles of the prototype 
cluster models (in the LMC tidal field) at 0, 30 and $300\,$Myr. 
The profiles are compared to two families of observationally determined density profiles: 
empirical King (1962) models based on Milky Way globular clusters and 
Elson, Fall \& Freeman (EFF: 1987) models based on young LMC clusters. 
The former are described by 
\begin{equation} 
\sigma \left( r \right) = \sigma_0 \left( 
\frac{1}{\left[ 1 + \left( r / r_{\rm c} \right)^2 \right]^{1/2}} - 
\frac{1}{\left[ 1 + \left( r_{\rm t} / r_{\rm c} \right)^2 \right]^{1/2}} \right)^2 \, , 
\end{equation} 
and the latter by 
\begin{equation} 
\sigma \left( r \right) = \sigma_0 \left( 1 + \frac{r^2}{a^2} \right)^{- \gamma / 2} \, , 
\end{equation} 
where 
\begin{equation} 
r_{\rm c} = a \left( 2^{2 / \gamma} - 1 \right)^{1/2} \, . 
\end{equation} 
The main difference between the model profiles is the introduction of $\gamma$ in the 
EFF profiles which allows greater flexibility to fit the slope of the distribution exterior 
to the core for clusters that do not show significant tidal truncation. 
As noted by Mackey \& Gilmore (2003) the two families overlap if we set $\gamma = 2$ 
in equation~(6) and assume $r_{\rm t} \rightarrow \infty$ in equation~(5). 
In fact, the profile of our prototype cluster at $30\,$Myr is best fitted by such 
a scenario: an EFF model with $r_{\rm c} = 0.28\,$pc, $\gamma = 2$ and a 
central surface density of $2\,500\, {\rm stars} \, {\rm pc}^{-2}$. 
The tidal radius at this time is $7\,$pc but using this in equation~(4) gives a profile 
that is too truncated to fit the data for $r > 1\,$pc. 
It is interesting to note that the functional form of the EFF profile was originally 
suggested by McGlynn (1984) to represent the equilibrium profile arising from an 
initially irregular distribution of objects but in the context of dissipationless collapse 
of proto-galaxies. 
The profile at $300\,$Myr is taken as representative of the density profile at late stages 
in the evolution of the bound cluster. 
At more advanced times the profile becomes progressively noisier as the number 
of bound stars decreases. 
This profile is well fitted by an EFF model with parameters of 
$r_{\rm c} = 0.16\,$pc, $\sigma_0 = 5\,500\, {\rm stars} \, {\rm pc}^{-2}$ 
and $\gamma = 2.6$, noting that this is the median slope determined by both 
EFF and Mackey \& Gilmore (2003) from samples of young and intermediate age LMC 
star clusters. 
However, the data at  $300\,$Myr are even better fitted by a King model with the 
same $r_{\rm c}$ and  $\sigma_0$ as well as the addition of $r_{\rm t} = 5\,$pc 
(see Figure~\ref{f:fig3}). 
This is to be expected as the model cluster orbiting close to the centre of the LMC 
is strongly truncated by this stage. 
In comparison, the density profiles of the cluster evolved in the standard MW tidal field 
are adequately fitted across the $30 - 100\,$Myr timeframe by King models with 
$r_{\rm c} = 0.26\,$pc and $\sigma_0 = 3\,500\, {\rm stars} \, {\rm pc}^{-2}$. 

We note that the prototype LMC cluster reaches a minimum in $r_{\rm c}$ at an age of 
approximately $400\,$Myr which is associated with the end of the core-collapse phase 
of evolution. 
The $r_{\rm c} / r_{\rm h}$ value at this point is 0.06 which is a typical core-collapse 
value for standard models of star cluster evolution (see Hurley 2007). 
There are two binaries in the cluster at this age, both residing in the core. 
The maximum number of binaries in the simulation is recorded shortly after the 
cluster forms ($5\,$Myr) when 36 binaries are present. 

\section{Summary}
\label{s:summ}

We have demonstrated the self-consistent formation of a bound star cluster 
that exhibits all the markings of a regular relaxed star cluster: 
dynamical equilibrium, core collapse, 
and an approximately spherical spatial distribution of the stars. 
This initially unrelaxed model, emerging from its progenitor giant molecular cloud, 
approaches virial equilibrium on the order of a few 
crossing times and before a half-mass relaxation time has passed 
-- a result that is free of the influence of the external tidal field. 
We have also found that the density profile of the model cluster resembles the 
distributions derived from observations 
-- King (1962) and Elson, Fall \& Freeman (1987) -- from the time that 
virial equilibrium is reached and onwards. 
Idealized models that start in virial equilibrium can comfortably adopt 
these density distributions 
when setting the initial positions of the stars. 
Our findings are distinct from previous simulations of the collapse of 
proto-globular clusters and proto-galaxies 
-- which also observed the emergence of a core-halo structure at equilibrium -- 
primarily in that the initial process of fragmentation is explicitly followed using a 
three-dimensional hydrodynamic scheme and the influence of a galaxy-scale 
tidal field is modelled throughout the process. 

This letter describes a model that demonstrates 
a working interface between 
simulations of star cluster formation and long-term star cluster dynamical evolution. 
In this model we have included several aspects, such as the use of equal-mass stars 
and a point-mass galaxy, that are commonly utilised in models of star cluster evolution 
but are not the most realistic approach. 
Future models will expand our study into a full investigation including: 
\begin{itemize}
\item the effect of a spectrum of stellar masses on the formation and 
         evolution of bound clusters -- associated processes such as stellar evolution 
         and mass segregation will affect the binding energy and the evolution timescales; 
\item a variety of cluster sizes, in terms of physical size and also a greater number of stars; 
\item an exploration of how the SFE assumed in the formation step affects the final 
          outcome, and also a greater understanding of the conditions required for 
          substantial binary formation; and, 
\item a variety of GMC locations within the host galaxy and 
          an extension to model the {\it live} gravitational potential of the host galaxy 
          in concert with the internal star cluster evolution, with dwarf, spiral and elliptical 
          hosts considered. 
\end{itemize}
This will allow a much more detailed study of the effects of internal and external 
dynamical processes on the formation and evolution of star clusters 
in a galactic context.

\section*{Acknowledgments}

K.B. acknowledges the financial support of the Australian Research Council. 
The GMC simulations 
in this work 
were performed on GRAPE systems  
kindly made available by the Center for Computational Astrophysics at the 
National Astronomical Observatory of Japan. 
We thank the anonymous referee for drawing important past works to our attention.

\label{lastpage}

\end{document}